\documentclass{PoS}

\usepackage{amssymb}
\usepackage{hepunits}
\usepackage{slashed}
\usepackage{wasysym}
\usepackage{graphicx}
\usepackage{subfig}

\newcommand{\beq}{\begin{equation}}
\newcommand{\eeq}{\end{equation}}
\newcommand{\bfig}{\begin{figure}[ht]\begin{center}}
\newcommand{\efig}{\end{center}\end{figure}}

\newcommand{\bv}[1]{\mathbf{#1}}
\newcommand{\hv}[1]{\hat{\mathbf{#1}}}

\newcommand{\Id}{\operatorname{Id}}
\newcommand{\Tr}{\operatorname{Tr}}

\newcommand{\msbar}{{\overline{\rm MS}}}

\newcommand{\rimom}{{\rm RI/MOM}}

\newcommand{\rismom}{{\rm RI/SMOM}}

\newcommand{\rismomqslash}{{\rismom_{\slashed{q}}}}
\newcommand{\rismomgamma}{{\rismom_{\gamma_{\mu}}}}

\newcommand{\tree}{{\rm tree}}
\newcommand{\lat}{{\rm lat}}

\newcommand{\Nf}{N_{\mathrm f}}

\title{NPR step-scaling across the charm threshold}

\ShortTitle{NPR step-scaling across the charm threshold}

\author{\speaker{Julien Frison}\\
        School of Physics and Astronomy, The University of Edinburgh, Edinburgh EH9 3JZ, UK\\
        E-mail: \email{jfrison@ph.ed.ac.uk}}

\author{Peter Boyle\\
        School of Physics and Astronomy, The University of Edinburgh, Edinburgh EH9 3JZ, UK\\
        E-mail: \email{pab@ph.ed.ac.uk}}

\author{Nicolas Garron\\
Department of Applied Mathematics \& Theoretical Physics,
University of Cambridge, Wilberforce Road, Cambridge CB3 0WA, United Kingdom\\
        E-mail: \email{ng389@damtp.cam.ac.uk}}

\author{RBC-UKQCD collaboration}

\abstract{Matching Non-Perturbative Renormalisation on the lattice and perturbative renormalisation would benefit from higher matching scales, which are needed for observables entering their per-mile era such as $B_K$. In this work we lay down a strategy, within the Rome-Southampton framework, to push this scale higher across the charm threshold, and apply it to an exploration of the $B_K$ running from $3\ \GeV$ to $9\ \GeV$. This is done on $\Nf=2+1+1$ ensembles generated by the RBC-UKQCD collaboration, and features a close study of the discretisation effects. }

\FullConference{The 32nd International Symposium on Lattice Field Theory\\
                 23-28 June, 2014\\
                 Columbia University New York, NY}

\begin{document}

\section{Introduction}

Many quantities in lattice field theory need to be renormalised in order to have meaningfull results in the continuum limit. This is in particular the case of the Kaon Bag Parameter $B_K$ which is here used as our sandbox, while our procedure can be applied to many other observables. Following the Rome-Southampton strategy, we perform this renormalisation non-perturbatively. However, this non-perturbative renormalisation (NPR) has at some point to be matched with the perturbative renormalisation used for Wilson coefficients and the whole phenomenology. The slow convergence of the pertubative series at lattice scales introduces large systematic errors, which are by far the dominant ones in the case of $B_K$.  

This work proposes a step-scaling strategy in dedicated ensembles to explore higher energies and therefore reduce the truncation errors. Since those energies lie beyond the charm mass we have to use $\Nf=2+1+1$ ensembles, and matching our $2+1$ low-energy results with the $2+1+1$ step-scaling requires a careful treatment of the charm threshold. One could fear that the slow running of $\alpha_s$ would make this approach hopeless at a fixed order of perturbation theory, unable to compete with the speed of improvement of the other errors. We will show that it is not the case.

Using the good statistical properties and low cost (compared to generating new ensembles) of our Green functions with four-volume sources, we also experiment new fitting strategies to try to extract as much information as possible from the various quantities we can build. As we probe energy ranges which have never been studied so far, we primarily focus on testing the potential of our new method designed for this new energy range. All the results given here, based on new M\"obius Domain Wall Fermion ensembles of the RBC-UKQCD collaboration, are still preliminary.

\section{RI/MOM schemes}

The $\rimom$ schemes are based on a simple renormalisation condition which is well-defined both on the lattice and in continuum perturbation theory: starting from an amputated Green function $\Gamma$ we define the renormalisation factor $Z$ such that we recover the tree-level expression through
\beq
Z\times\left.\Tr(P\Gamma)\right|_{\lat} = \left.\Tr(P\Gamma)\right|_{\tree} .
\eeq
The projectors $P$ can be chosen arbitrarily and contribute to the unique definition of the scheme, while the momenta flowing into $\Gamma$ also set a renormalisation scale.

In particular we will use the $\rismomgamma$ and $\rismomqslash$ schemes which are defined by the momentum pattern of Fig.~\ref{fig:RISMOM-tree} and the following projectors for $2$-point and $4$-point functions: 
\begin{eqnarray}
P^{A,\gamma}_\mu &=&  \Id_{\mathrm{color}}\otimes(\gamma_\mu\gamma_5) \quad\mathrm{and}\quad
P^{A,\slashed{q}} =  \Id_{\mathrm{color}}\otimes(\slashed{q}\gamma_5)/q^2\\
P^{VV+AA,\gamma}_\mu &=&  \left(\Id\otimes\Id\right)_{\mathrm{color}}\otimes\left[\gamma_\mu\otimes\gamma_\mu+(\gamma_\mu\gamma_5)\otimes(\gamma_\mu\gamma_5)\right]\\
P^{VV+AA,\slashed{q}} &=& \left(\Id\otimes\Id\right)_{\mathrm{color}}\otimes\left[\slashed{q}\otimes\slashed{q}+(\slashed{q}\gamma_5)\otimes(\slashed{q}\gamma_5)\right]/q^2
\end{eqnarray}

In the case of $B_K$ we are interested in the ratio
$Z_{B_K} = Z_{VV+AA}/Z_A^2$, 
which combines the renormalisation factors of the four-quark and two-quarks operators involved in this $B_K$ ratio \cite{Aoki:2007xm}. 

\bfig
	\hfill
  \includegraphics[width=0.33\linewidth]{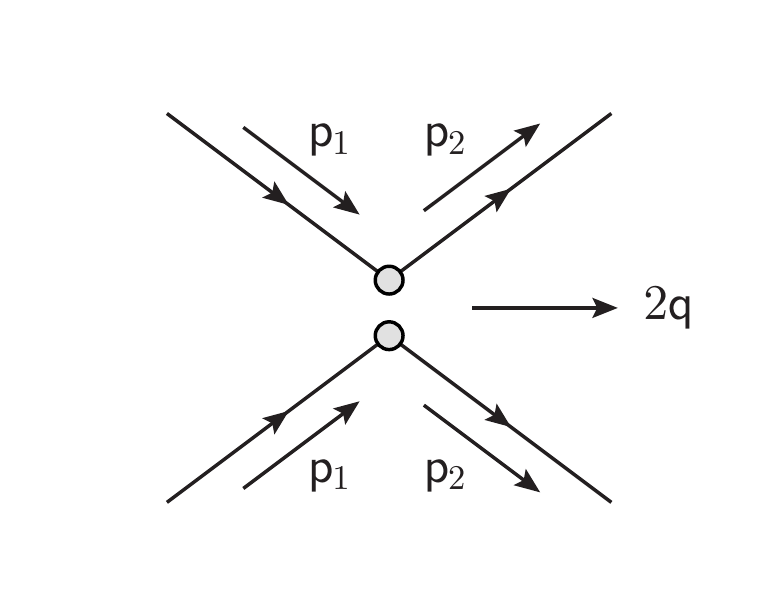}
	\hfill
  \includegraphics[width=0.33\linewidth]{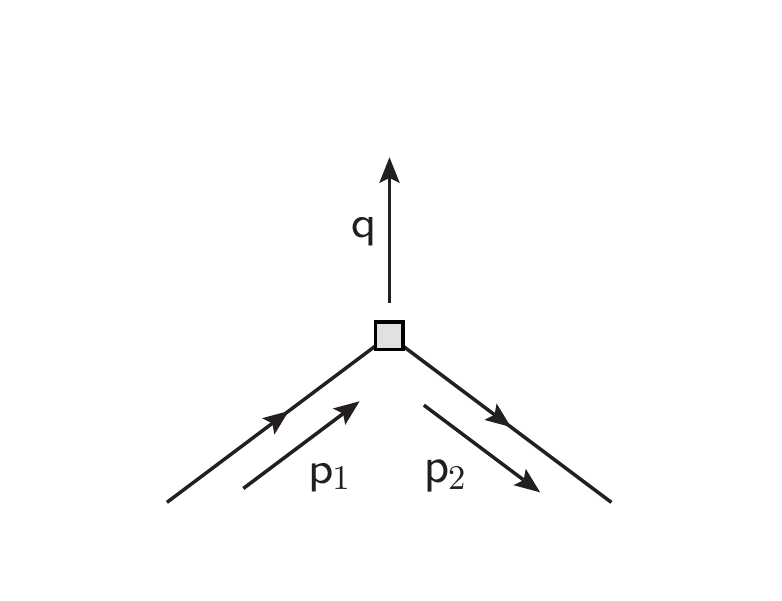}
        \hspace{1.5cm} 
	\hfill
  \caption{In symmetric non-exceptional schemes, momenta are chosen such that $p_1^2=p_2^2=(p_1-p_2)^2$, which leads to some momentum $q$ (or $2q$) flowing through the operator and ensures that no partial sum of the momenta cancels out. The chiral properties and the convergence of the perturbative series are therefore improved. }
  \label{fig:RISMOM-tree}
\efig

Once this Non-Perturbative Renormalisation has been performed, one can compute the continuum limit of the renormalised matrix elements in the $\rismom$ schemes. It is then usually converted to $\msbar$ at the end of the process, which introduces some perturbative errors but does not need us to worry about lattice perturbation theory. 
Alternatively, $B_K$ is often expressed in a Renormalisation Group Invariant (RGI) form which absorbs the $1$-loop perturbative running:
\beq
\hat B_K = \left(\frac{g(\mu)^2}{4\pi}\right)^{-\gamma_0/(2\beta_0)}\times\exp\left\{\int_0^{g(\mu)}dg\left(\frac{\gamma(g)}{\beta(g)}+\frac{\gamma_0}{\beta_0g}\right)\right\}B_K(\mu)
\eeq

\section{Step-scaling and parametrisation of $Z$}

While those $Z$ depend on all the details of each lattice (choice of action, breaking of hypercubic symmetry, \dots), the step-scaling function\cite{Arthur:2010ht} has an universal limit in the continuum. Up to discretisation effects included in $d_a\sim 1$,
\begin{equation}
\sigma_a(p,p_0) \equiv \frac{Z_a(p)}{Z_a(p_0)} \sim \sigma_0(p,p_0) d_a(p,p_0) .
\end{equation}

In this work we explore the possibilities of parametrisation of $d_a$ in order to keep those discretisation errors under control to larger scales. This approach has already been successfully followed in \cite{Constantinou:2013ada} based on lattice perturbation theory but here we try a simpler and more general method. We can indeed notice that, in the chiral limit and once we've fixed the momenta orientations, the perturbative expressions has to be of the form
\begin{equation}
Z_a(p)^{-1} = 1 + g_b^2 \left[ F^{\mathrm{cont}}(ap) + (ap)^2F^{\mathrm{discr}}(ap) + O(a^2\log^ka)\right] +g_b^4 G(ap) + \cdots
\label{eq:Z_pert}
\end{equation}
Replacing the bare coupling $g_b$ by a boosted coupling
\beq
g^{*2} = \frac{g_b^2}{\frac{1}{3}\langle\Tr U_{\mathrm{plaq}}\rangle} \left(1+cg^{*2}+ O(g^{*4}) \right) ,
\eeq
which has negligible discretisation error as long as it is associated with scale-setting from a low-energy observable, 
we see that most of the discretisation terms should depend on $ap$ for hard on-shell amplitudes ($p$ becomes the only physically relevant scale). The $g^*$ running only brings a logarithmically slow dependence on $a$. This justifies our hope to find information about the discretisation effects in the $p$ dependence of each $Z_a$. At lower scales 
the series blows up
, while in practice $Z(p)$ acquires a strong curvature. But for reasonably large scales we expect to be able to Taylor expand the discretisation effect 
in terms of a finite number of fit parameters: 
\begin{equation}
d_a(p) \sim 1+\alpha(ap)^2+\beta(a)^2+\gamma(ap)^4+\delta(ap)^2(a)^2 .
\end{equation}
The addition of terms in $a$ instead of $ap$ makes this expansion even more general than Eq.~\ref{eq:Z_pert}, and can have as main origins non-zero mass and non-pertubative effects, which are expected to be rather small. The dominant term should be $\alpha(ap)^2$ when we push to moderately high scales. 

Those parameters describing $d_a$ are actually direction-dependent because of symmetry breaking. It is then interesting, instead of fixing the direction once for all, to look at those ratios between momenta of different directions (but same norm):
\begin{equation}
\sigma_a(\bv p,\bv{p'}) = \frac{\sigma_a(\bv p,\bv{p_1})}{\sigma_a(\bv{p'},\bv{p_1})} \sim \frac{d_a^{\hv{p}}(p,p_1)}{d_a^{\hv{p'}}(p',p_1)} .
\end{equation}
In particular, in the continuum the Lorentz symmetry is restored and this ratio goes to $1$. Any fitting model or result incompatible with this universality constraint would have to be discarded. 

Finally, let us now look at something similar when we combine two different lattices: 
\begin{equation}
\frac{Z_a(\bv p)}{Z_{a'}(\bv{p})} = \frac{Z_a(\bv {p_0})s_a(\bv p,\bv{p_1})}{Z_{a'}(\bv {p_0})s_{a'}(\bv{p},\bv{p_1})} \sim \frac{Z_a(\bv {p_0})}{Z_{a'}(\bv {p_0})} \times \frac{d_a^{\hv{p}}(p,p_1)}{d_{a'}^{\hv{p}}(p,p_1)} .
\end{equation}
Once again the running has been eliminated and this ratio only consists of discretisation errors, up to a constant. This contains the information we would use if we just performed one separate continuum extrapolation for every $p$. It has a straightforward interpretation but having enough lattice spacings to check our scaling is much more expensive than having different momenta, and moreover its signal gets spoiled by relatively large uncertainties, mainly from scale-setting.

Our strategy will consist in adding in a global fit a large selection of those different kinds of ratios, so that we combine their advantages, both in terms of numerical signal and in terms of the number of constraints to test our model. 

\section{Strategy for the charm quark}

The $\rismom$ are massless schemes and the conversion factors are only known in such a framework. However, in practice both the lattice and the physical world have non-zero masses. For the light quark this just gets solved by extrapolating our lattice results to the chiral limit. For the charm quark we have to make sure that all the limits are taken in the correct order, since the charm threshold is in the middle of our Rome-Southampton window and the intermediate massive scheme goes from an effective $\Nf=2+1$ massless theory to an effective $\Nf=2+1+1$ massless theory.

For an operator $\cal O$ whose matrix element has been determined on a $\Nf=2+1$ lattice and normalised at a scale $\mu_1$, we can write
\begin{eqnarray}
{\cal O}_{\Nf=2+1}(\mu_1) &&= \frac{Z_{\Nf=2+1}(\mu_1)}{Z_{\Nf=2+1+1}(\mu_1;m_c=m_c^{\mathrm{phys}})} \frac{Z_{\Nf=2+1+1}(\mu_1;m_c=m_c^{\mathrm{phys}})}{Z_{\Nf=2+1+1}(\mu_1;m_c=m_0)} \frac{Z_{\Nf=2+1+1}(\mu_1;m_c=m_0)}{Z_{\Nf=2+1+1}(\mu_2;m_c=m_0)}\times\nonumber\\
 &&\frac{Z_{\Nf=2+1+1}(\mu_2;m_c=m_0)}{Z_{\Nf=2+1+1}(\mu_2;m_c=0)}  \frac{Z_{\Nf=2+1+1}(\mu_2;m_c=0)}{Z_{\Nf=2+1+1}(M_W;m_c=0)} {\cal O}_{\Nf=2+1+1}(M_W) .
\end{eqnarray}
The $\mu_2\to M_W$ step-scaling at $m_c=0$ is done perturbatively, so that the scale $\mu_2$ determines the size of the truncation error. The other step-scaling, $\mu_1\to\mu_2$, is done non-perturbatively at $m_c=m_0$. One obvious choice is $m_0=m_c^{\mathrm{phys}}$, in which case all the other ratios tend towards $1$ as long as $\mu_1\ll m_c^{\mathrm{phys}}\ll\mu_2$. Another choice is the extrapolation to $m_0=0$, which leaves $Z_{\Nf=2+1+1}(\mu_1;m_c=m_c^{\mathrm{phys}}) / Z_{\Nf=2+1+1}(\mu_1;m_c\to 0)$ as a non-vanishing number which has to be determined, but only requires $\mu_1\ll m_c^{\mathrm{phys}}$ while keeping $\mu_2$ absolutely arbitrary.

Those preliminary results will be based on the $m_0=m_c^{\mathrm{phys}}$ approach. However, we do not yet have enough data to ensure exactly $m_c=m_c^{\mathrm{phys}}$, nor that $\mu_1$ is small enough to be immune to small changes of $m_c$, and we used our single light-$m_c$ ensemble to bound the systematics introduced: 
\begin{equation}
\left\|1-\frac{Z_{\Nf=2+1+1}^{\beta=5.70}(\mu_1;m_c=0.243)}{Z_{\Nf=2+1+1}^{\beta=5.70}(\mu_1;m_c=0.1)}\right\|_{\infty}\lesssim 2\permil
\end{equation}

\section{Results}

We present our $\Nf=2+1+1$ ensembles in Tab.~\ref{tab:ensembles}. Amongst the three ensembles included in our global fit, two of them have Green functions computed for orientations I and II and the $\beta=5.77$ only has orientation I, with those orientations being: 
\begin{eqnarray}
\mathrm{I}: && p_1\propto [0,1,1,0] ,\  p_2\propto [-1,0,1,0],\  p_1-p_2\propto [1,1,0,0]\\
\mathrm{II}: && p_1\propto [1,1,1,1] ,\  p_2\propto [1,1,1,-1],\  p_1-p_2\propto [0,0,0,2]. 
\end{eqnarray} 
We consider all the possible ratios between those ensembles and orientations, which are ploted in Fig.~\ref{fig:fit_ratios}. For ratios of different ensemble we have to slightly interpolate to common physical $p^2$, which degrades the error bars. The ratios of two different orientations, on the other hand, show very tiny errors bars and let us clearly see the $p^2$ dependence of those discretisation effects, which is almost entirely an $(ap)^2$ dependence. Before even trying the global fit we can check that at a given $a^{-1}$ the discretisation errors are well described by a linear term in $(ap)^2$ until $(ap)^2\sim 1.5$, and by parabolas including $(ap)^4$ terms until $(ap)^2\sim 3$. For $(ap)^2>3$ the discretisation errors explode (compared to our tiny error bars) and no higher monomials can save the fits, we therefore cut the data to this point before sending it to the global fit. For our finest ensemble it corresponds to a cut at $7.4\ \GeV$, while limiting ourself to $(ap)^2$ terms would led us around $5\ \GeV$, in coherence with the typical values of $(ap)^2$ used in the past for coarser lattices \cite{Frison:2013fga}.

The global fit shows satisfying $\chi^2\lesssim 1$ for both schemes. Once we use the resulting fit parameters to subtract the discretisation error of the step-scaling (Fig.~\ref{fig:BK_RGI_corr}), the absence of residual dependence on the lattice spacing and the momentum orientation, surprisingly even up to $9\ \GeV$, is another evidence of the success of our fit. Additionally, this subtraction brings the $\rismomqslash$ step-scaling much closer to its $1$-loop expression, 
with the difference being barely significant. 

\begin{table}
\centering
  \begin{tabular}{|c|c|c|c|c|c|c|}
    \hline
    $\beta$ & $L^3\times T\times L_5$ & $m_l$ & $m_c$ & $a^{-1}$ & $(ap)^2\sim 1.5$ & $(ap)^2\sim 3$\\
    \hline
    $5.70$ & $32^3\times 64\times 12$ & $\mathbf{0.0047}$ & $\mathbf{0.243}$ & $3.0\ \GeV$ & $3.7\ \GeV$ & $5.2\ \GeV$ \\
    $5.70$ & $32^3\times 64\times 12$ & $0.002$ & $0.243$ & $3.0\ \GeV$ & $3.7\ \GeV$ & $5.2\ \GeV$\\
    $5.70$ & $32^3\times 64\times 12$ & $0.0047$ & $0.1$ & $3.0\ \GeV$ & $3.7\ \GeV$ & $5.2\ \GeV$\\
    $5.77$ & $32^3\times 64\times 12$ & $\mathbf{0.0044}$ & $\mathbf{0.213}$ & $3.6\ \GeV$ & $4.4\ \GeV$ & $6.2\ \GeV$\\
    $5.84$ & $32^3\times 64\times 12$ & $\mathbf{0.0041}$ & $\mathbf{0.183}$ & $4.3\ \GeV$ & $5.3\ \GeV$ & $7.4\ \GeV$\\
    $5.84$ & $32^3\times 64\times 12$ & $0.002$ & $0.183$ & $4.3\ \GeV$ & $5.3\ \GeV$ & $7.4\ \GeV$\\
    \hline
  \end{tabular}
  \caption{$\Nf=2+1+1$ ensembles used in this analysis. We put in boldface the masses of the main ensembles, which are included in the global fit. The role of the other ensembles is currently limited to estimating the systematics, while we eventually want a massless extrapolation. The values of lattice spacings, obtained through Wilson flow, are still preliminary. In the last two columns we translate in physical units the typical cut-off which has shown to be necessary for order $a^2$ (resp. $a^4$) fits to only introduce substatistical systematics. }
  \label{tab:ensembles}
\end{table}

\bfig
	\hfill
  \includegraphics[width=0.45\linewidth]{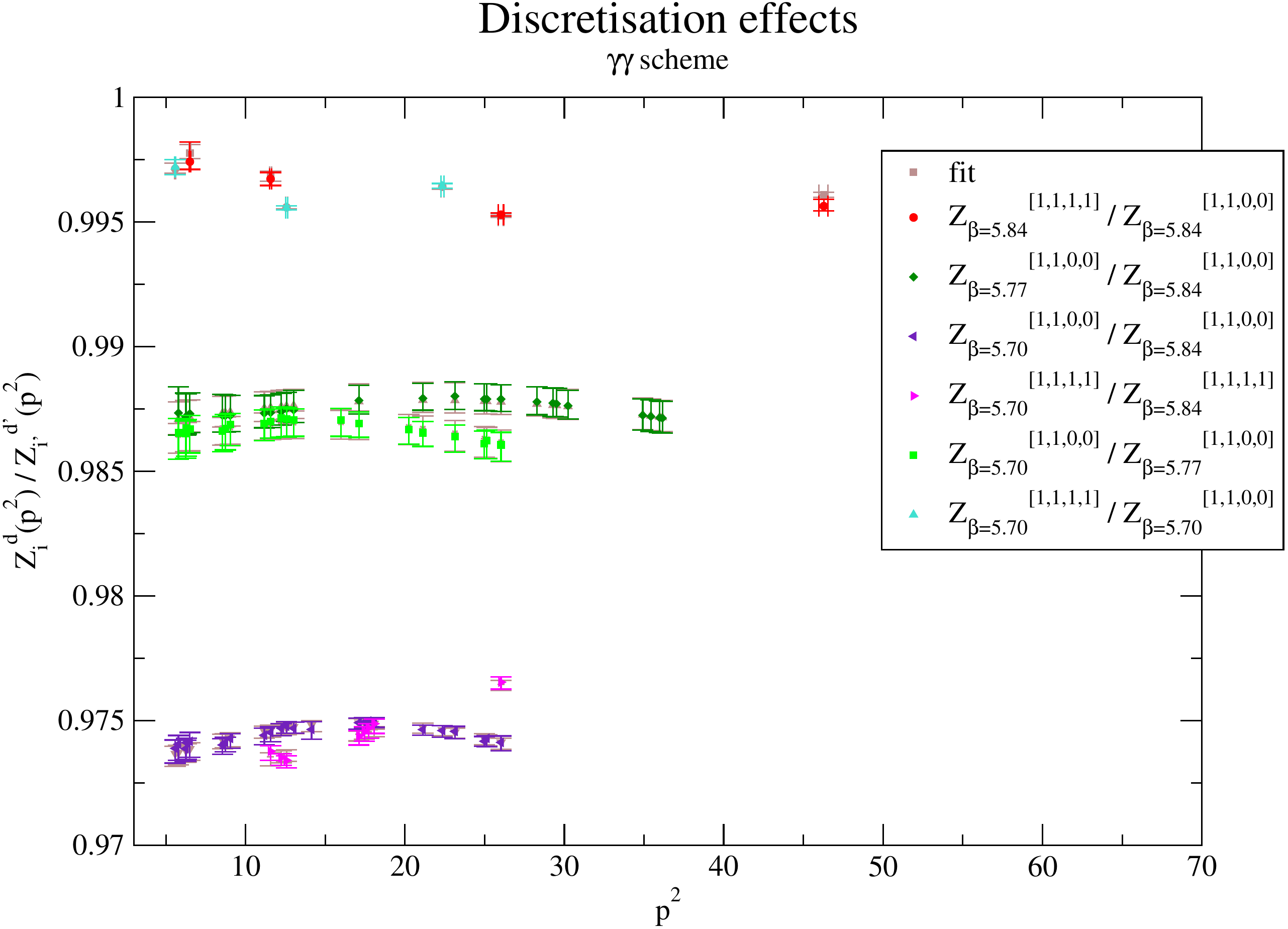}
	\hfill
  \includegraphics[width=0.45\linewidth]{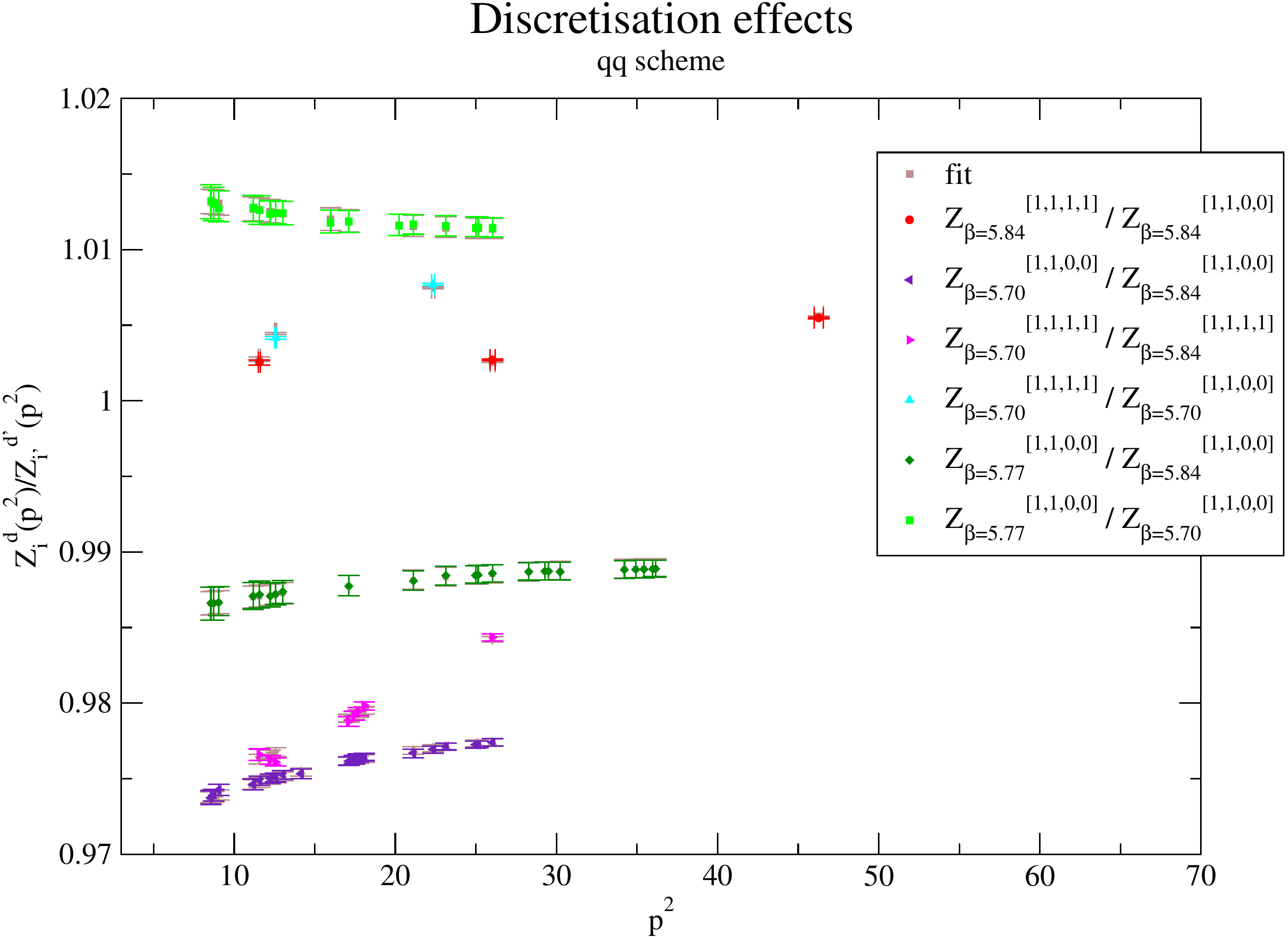}
	\hfill
  \caption{All those ratios should be constant up to discretisation effects. The data is so close to the fit result (brown) that the latter is barely visible. One can see the curvature on several curves at $(ap)^2\sim 1.5$, which gives us sensitivity to the $(ap)^4$ terms which we may begin to include in our fits, particularly in $\rismomgamma$. }
  \label{fig:fit_ratios}
\efig

\bfig
\subfloat{\includegraphics[width=0.6\linewidth]{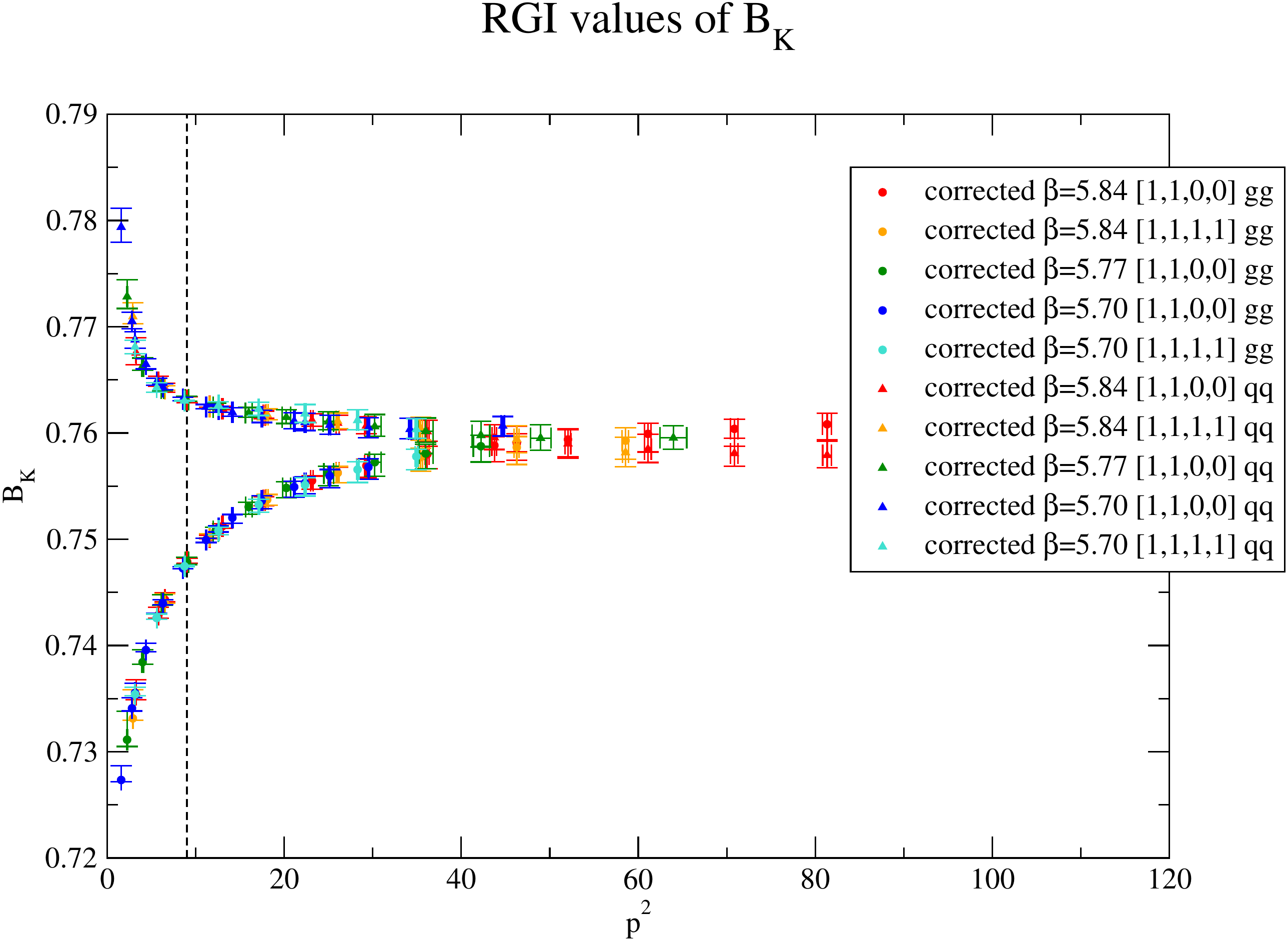} }
\subfloat{\raisebox{3cm}{
\begin{tabular}{|c|c|c|}
\hline
$p$ & $\hat B_K^\gamma$ & $\hat B_K^{\slashed{q}}$\\
\hline
$3\ \GeV$ & $0.7478$ & $0.7631$ \cite{Blum:2014}\\
$4\ \GeV$ & $0.753(1)$ & $0.762(1)$ \\
$5\ \GeV$ & $0.756(1)$ & $0.761(1)$ \\
$6\ \GeV$ & $0.758(1)$ & $0.760(1)$ \\
$7\ \GeV$ & $0.759(2)$ & $0.759(2)$ \\
$9\ \GeV$ & $0.761(2)$ & $0.758(2)$ \\
\hline
\end{tabular}
}}
  \caption{We plot the RGI 4-flavour running as a preliminary demonstration of our method. While discretisation effects are importants at the scale of this plot (up to 2\%), their subtraction leads to a flat $\hat B_K^{\slashed{q}}$ and a perfect convergence between the two intermediate schemes, within our error bars (systematic errors of up to a few per-mile have to been added to these preliminary numbers). Note that the $p^2>50\ \GeV^2$ region of this plot is only shown for the curiosity of the reader, and is not included in the global fit. An other point worth noting is that the four-flavour RGI does {\it not} theoretically have to converge with the three-flavour RGI in any way.}
  \label{fig:BK_RGI_corr}
\efig

\section{Conclusion}

Using both finer lattices and a more involved analysis, we have been able to raise the scale from $3\ \GeV$ to $5\ \GeV$ and explore even higher. It appears $6-7\ \GeV$ is sufficient to bring the two intermediate schemes into statistical agreement and thus vastly reduce the perturbative error. It also led to interesting comparison of the runnings between our two scheme: while they finally converge, the convergence is much faster for $\rismomqslash$. This is therefore bringing us new knowledge to better estimate the truncation error of our previous works at $3\ \GeV$, which up to now was strongly overestimated, even though that is not something we could have assumed a priori before having this new data at higher scales. Finally, 
our preliminary results are supported both by the convergence of the two schemes and by the several constraints included in the global fit, paving the way for a much stronger check that what has been done in the past at a fixed momentum. 

However, several sources of systematics are not fully under control in those preliminary results, each contributing to the per-mile level. The main one is probably the dependence on $m_c$, where new data has to be generated. Then the scale-setting could be changed by a non-negligible overall factor as we have not yet corrected the $m_l$ dependence of the continuum limit of $w_0$. Finally, the finite volume effects are expected to be negligible but we do not have an indisputable proof.

The physical reach of this $B_K$ preliminary result is actually limited by the fact the error on $B_K$ is already much smaller than on the other experimental and phenomenological inputs of $\epsilon_K$. But the quality of $B_K$ could also be seen as encouragement to tackle those new issues, and renormalisation will continue to be a question of fundamental interest.  Additionally this strategy can be applied to many other operators, in particular other four-quark operators entering in ``SUSY $B_K$'' and $K\to\pi\pi$. 

\section{Acknowledgements}
N.G. acknowledges support from STFC under the grand ST/J000434/1
and from the European Union under Grant Agreement number 238353 (ITN STRONGnet).

\end{document}